\documentclass[aps,prl,showpacs,twocolumn]{revtex4-2}%
\usepackage{graphicx}
\usepackage{graphicx}
\usepackage{amsmath}
\usepackage{amsfonts}
\usepackage{xcolor}
\usepackage[T2A]{fontenc}
\usepackage{braket}

\providecommand{\LyX}{L\kern-.1667em\lower.25em\hbox{Y}\kern-.125emX\@}

\begin{document}
	\title{Heterogeneity Induces Cyclops States in Kuramoto Networks with Higher-Mode Coupling}
	
	\author{Maxim I. Bolotov$^{1}$, Lev A. Smirnov$^{1}$, Vyacheslav O. Munyayev$^{1}$, Grigory V. Osipov$^{1}$, and Igor Belykh$^{2}$\footnote{Corresponding author, e-mail: ibelykh@gsu.edu}}
	
	\address{$^1$Department of Control Theory, Lobachevsky State University of Nizhny Novgorod,
		23 Gagarin Avenue, Nizhny Novgorod, 603022, Russia\\
		$^2$Department of Mathematics and Statistics and Neuroscience Institute, Georgia State University, P.O. Box 4110, Atlanta, Georgia, 30302-410, USA}	
	\begin{abstract}
		Disorder is often seen as detrimental to collective dynamics, yet recent work has shown that heterogeneity can enhance network synchronization. However, its constructive role in stabilizing nontrivial cooperative patterns remains largely unexplored. In this Letter, we show that frequency heterogeneity among oscillators can induce stable Cyclops and cluster states in Kuramoto networks with higher-mode coupling, even though these states are unstable in the identical oscillator case. Cyclops states, introduced in [Munyaev et al., Phys. Rev. Lett. 130, 107021 (2023)], feature two synchronized clusters and a solitary oscillator, requiring a delicate phase balance. Surprisingly, heterogeneity alone is sufficient to stabilize these patterns across a broad range of detuning values without needing to be compensated by other forms of disorder or external tuning.  We introduce a mesoscopic collective coordinate approach that connects microscopic frequency structure, captured by the solitary oscillator’s influence, with mean-field cluster-level stability. This constructive approach identifies favorable ranges of heterogeneity and suitable initial conditions for inducing robust multi-state dynamics, offering a foundation for their analysis in broader classes of heterogeneous biological and engineering networks.
	\end{abstract}
	\date{\today}
	\draft \maketitle
	
	\textit{Introduction.} 
	Networks of phase oscillators serve as a canonical framework for studying emergent collective behavior in complex systems \cite{hoppensteadt2012weakly,tinsley2012chimera,motter2013spontaneous}. Among these, the Kuramoto model of coupled phase oscillators in one and two dimensions \cite{kuramoto1975self,strogatz2000kuramoto,ermentrout} has become a central model for exploring synchronization and related phenomena \cite{acebron,ott2008low}. It supports a rich spectrum of dynamics—including full \cite{tanaka1997first,tanaka1997self,ji2014low,munyaev2020analytical,komarov2014synchronization}, partial \cite{martens2009exact,barabash2021partial}, explosive \cite{gomez2011explosive,ji2013cluster,skardal2014disorder}, chimera \cite{kuramoto2002coexistence,abrams2004chimera,abrams2008solvable,panaggio2015chimera,bolotov2016marginal,bolotov2018simple}, solitary \cite{jaros2015chimera,maistrenko2017smallest,jaros2018solitary,teichmann2019solitary,munyayev2022stability}, cluster \cite{belykh2016bistability,brister2020three,ronge2021splay,zhang2020symmetry}, generalized splay \cite{berner2021generalized}, and cyclops states \cite{munyayev2023cyclops,bolotov2024breathing}.
	Historically, synchronization in such systems has been studied predominantly under the assumption of identical or nearly identical oscillators, with heterogeneity and noise treated as destabilizing perturbations. However, a growing body of research has demonstrated that disorder can, counterintuitively, promote coherence. It has been shown that disorder in oscillator frequencies, delays, coupling strengths, or noise can suppress spatiotemporal chaos and induce synchronization in diverse physical, biological, and engineered systems \cite{braiman1995taming,braiman1995disorder,qi2003ordering,corral1997self,neiman1999noise,alexeeva2000impurity,gavrielides1998spatiotemporal,braiman1999tuning,brandt2006synchronization,montaseri2016diversity,li2012parameter,tessone2006diversity,skardal2014disorder,taylor2016synchronization,nishikawa2016symmetric,belykh2017foot,hart2019topological,nicolaou2019multifaceted,nicolaou2020coherent,molnar2020network,zhang2021random,molnar2021asymmetry,punetha2022heterogeneity,sugitani2021synchronizing,eliezer2022controlling,honari2025spectral}. For instance, while frequency heterogeneity often degrades synchrony in laser oscillator arrays, introducing additional disorder in delay times has been shown to reverse this effect and restore near-perfect phase synchronization \cite{nair2021using}.
	A particularly compelling concept is converse symmetry breaking \cite{nishikawa2016symmetric,nicolaou2019multifaceted,zhang2021random,molnar2020network,molnar2021asymmetry}, where the introduction of disorder enables synchronous states that are inaccessible in fully symmetric systems. For example, structural heterogeneity in the coupling topology can stabilize complete synchronization \cite{hart2019topological}, and carefully tuned frequency distributions can optimize synchrony in otherwise unfavorable networks \cite{taylor2016synchronization}. In adaptively coupled systems, the interaction of multiple disorder types, such as nodal heterogeneity and adaptive weights, can induce two distinct non-equilibrium phase transitions en route to synchrony \cite{fialkowski2023heterogeneous}. Despite these advances, much of the prior work has focused on disorder-induced global synchronization. The ability of disorder to induce complex cooperative states, such as multi-cluster or other higher-dimensional cooperative dynamics, remains poorly understood, especially when such states must persist under changes in the coupling strengths or distribution of heterogeneity and without relying on additional sources of disorder or parameter fine-tuning.
	
	In this Letter, we address this conceptual knowledge gap and demonstrate that oscillator frequency heterogeneity can robustly stabilize cyclops and cluster states in the second-order Kuramoto model with higher-harmonic coupling, even though these states are unstable in networks of identical oscillators. In our previous work \cite{munyayev2023cyclops}, we introduced cyclops states, characterized by two coherent clusters and a solitary oscillator with the relative phase positioned between the clusters, evocative of the Cyclops' eye. There, we showed that higher-harmonic coupling can make these states global attractors in repulsively coupled Kuramoto networks and theta-neuron models with adaptive coupling. Here, we report a counterintuitive finding that simple, non-engineered frequency heterogeneity, such as a uniform distribution, is sufficient to stabilize cyclops and related cluster states. These patterns emerge from a wide set of initial conditions and remain stable over a broad range of oscillator heterogeneity. To explain this disorder-induced stabilization, we introduce a mesoscopic reduction based on the collective coordinate framework \cite{cox2006bistable,gottwald2015model,hancock2018model,smith2019chaos,berner2023synchronization,smith2020model,yadav2024disparity,fialkowski2023heterogeneous}. Our approach captures the interplay between microscopic frequency heterogeneity and macroscopic cluster dynamics and adopts the linear ansatz \cite{smith2020model} to tracks the relative dynamics between synchronized clusters and a solitary oscillator. This formulation is particularly suited to multi-cluster patterns like cyclops states and networks with higher-mode coupling. It reveals the surprising role of disorder in transforming unstable, seemingly fragile configurations, requiring a delicate phase balance, into robust and prevalent regimes, due to a nontrivial interplay between the second-order parameter and the second coupling harmonic. Moreover, it provides a constructive means to identify favorable heterogeneity ranges and initial conditions that support such complex multi-state dynamics.
	
	\textit{The network model.} 
	We study a network of second-order Kuramoto–Sakaguchi oscillators with two-harmonic coupling 
	\cite{munyayev2023cyclops,bolotov2024breathing}:\vspace{-3mm}
	\begin{equation}
		m \ddot\theta_n + \dot\theta_n = \omega_n + \sum_{k=1}^N \sum_{q=1}^{2} \frac{\varepsilon_q}{N} \sin\big(q(\theta_k - \theta_n) - \alpha_q\big), \label{eq:system_origin}
		\vspace{-3mm}
	\end{equation}
	where $\theta_n \in [-\pi, \pi)$ is the phase of the $n$th oscillator, and $n = 1, \dots, N$, with 
	$N$ assumed to be odd throughout the paper.
	The coupling function $H(\theta_k - \theta_n)$ includes first and second harmonics with strengths $\varepsilon_1$, $\varepsilon_2$ and phase lags $\alpha_{1,2} \in [0, \pi/2)$, chosen to ensure attractive interactions with $H'(0) > 0$. Inertia is fixed at $m = 1$ to enable breathing cluster dynamics \cite{belykh2016bistability}, which are absent in the first-order model. To highlight the generality of the disorder-induced effects, natural frequencies $\omega_n$ are drawn from a uniform distribution on $[\omega_0 - \nu, \omega_0 + \nu]$, where 
	$\nu$ is the half-width of the distribution. No fine-tuning or engineered frequency profile is required; we take $\omega_0 = 0$ without loss of generality.
	\begin{figure}[t!]\label{fig1}
		\center
		\includegraphics[width=0.45\textwidth]{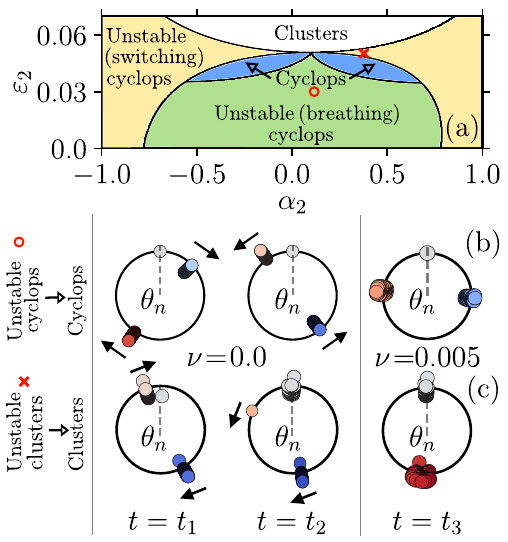} 
		\caption{Heterogeneity-induced stabilization. (a) Stability diagram for identical oscillators ($\omega_n = \text{const}$) showing stable (blue), unstable (breathing) (green), and unstable (switching) (yellow) cyclops, and stable two-cluster states (white). Red circle ($\alpha_2 = 0.1$, $\varepsilon_2 = 0.03$) and cross ($\alpha_2=0.35$, $\varepsilon_2=0.05$) mark two representative points for unstable cyclops and cluster regimes, respectively. (b,c) Phase snapshots (in the solitary-anchored rotating frame) before and after introducing frequency heterogeneity ($\nu = 0.005$). Arrows indicate cluster oscillations; color encodes phase difference from the solitary oscillator. (b) Stabilization of unstable cyclops states at $\alpha_2 = 0.1$, $\varepsilon_2 = 0.03$. (c) Stabilization of unstable two-cluster state at $\alpha_2 = 0.35$, $\varepsilon_2 = 0.05$. Other parameters: $N = 101$, $\alpha_1 = 1.57$, $\varepsilon_1 = 1.0$.}
		\vspace{-7mm}
	\end{figure}
	
	For identical oscillators with $\omega_1=\cdots=\omega_N = \omega,$  the system~\eqref{eq:system_origin} supports cyclops states, in which the population splits into two symmetric clusters of size $K = (N-1)/2$ and a solitary oscillator. Their phases take the form $\theta_1 = \cdots = \theta_K = x + \Omega t$, $\theta_{K+1} = \cdots = \theta_{2K} = y + \Omega t$, and $\theta_N = \Omega t$, where $x$ and $y$ are constant phase differences relative to the $N$th solitary oscillator. These states constitute a particular class of generalized splay states, characterized by a vanishing first-order Kuramoto order parameter $R_1\!\left(t\right)=\!\sum\limits_{n=1}^N\!{e^{i\theta_{n}}}\big/N=r_1 e^{i\Theta_1},$ with 
	$r_1=0$ and a nonzero second-order parameter  $	R_2\!\left(t\right)=\!\sum\limits_{n=1}^N\!{e^{i 2\theta_{n}}}\big/N=r_2 e^{i\Theta_2},$ where 
	$r_2$ governs their stability and measures the degree of cluster synchrony \cite{berner2021generalized,skardal2011cluster,munyayev2023cyclops}. Depending on the coupling strengths $\varepsilon_1, \varepsilon_2$ and phase lags $\alpha_1, \alpha_2$, the system~\eqref{eq:system_origin} may have
	up to $16$ distinct stationary cyclops state—with different ordered pairs $(x, y)$. The permissible values of the phase differences $x$ and $y$ and collective rotation frequency $\Omega$ were computed analytically in \cite{bolotov2024breathing}. 
	As shown in \cite{munyayev2023cyclops}, the presence of second-harmonic coupling promotes cyclops states as dominant rhythms in repulsive networks and as strong attractors coexisting with global synchrony under attractive coupling. These states can lose stability via two distinct bifurcation scenarios. In the first, an Andronov–Hopf bifurcation destabilizes the stationary phase differences while preserving the intra-cluster structure, giving rise to breathing cyclops states with periodically oscillating inter-cluster phase differences, $x(t)$ and $y(t)$. In the second scenario, structural instability leads either to asymmetric two-cluster states or rapid reconfiguration into switching cyclops states, where the solitary oscillator and cluster composition change recurrently. This dynamic parallels the death–birth cycling observed in blinking chimeras \cite{goldschmidt2019blinking}, in which coherent structures disintegrate and reassemble in new forms.
	
	Figure~1a shows the stability diagram for cyclops states in the identical oscillator case, highlighting the boundaries associated with the two bifurcation scenarios. The white region corresponds to stable two-cluster states of the form $\theta_1 = \cdots = \theta_K = \Theta_1 + \Omega t$, $\theta_{K+1} = \cdots = \theta_N = \Theta_2 + \Omega t$. This configuration can be viewed as a degenerate cyclops state with $y = 0$. However, within the cyclops stability region (blue), these two-cluster states become unstable. We select two representative points in the stability diagram (Fig.~1a): a red circle marking an unstable cyclops state and a red cross marking an unstable two-cluster state. In the following, we demonstrate that introducing heterogeneity in the oscillators' natural frequencies can stabilize both configurations.
	
	In the presence of heterogeneous $\omega_n$, the perfectly symmetric cyclops and two-cluster states of the identical oscillator case no longer exist. However, their analogs can still be defined. In such a non-identical oscillator cyclops state, oscillator phases within each cluster are no longer identical but remain tightly grouped. Specifically, we write $\theta_n(t) = \varphi_n + \theta_N(t)$, where $|\varphi_j - \varphi_k| = C_{jk} < \bar{\phi}$ for $j, k \in \{1, \dots, K\}$ and all $j, k \in \{K+1, \dots, 2K\}$, respectively.  The solitary oscillator's relative phase is $\varphi_N=0.$ Similarly, a two-cluster state satisfies $\theta_n(t) = \varphi_n + \Omega t$, with $|\varphi_j - \varphi_k| = C_{jk}$ for indices within the same cluster. While the constants $C_{jk}$ depend on the frequency width $\nu$, they must remain bounded to prevent phase slips. \vspace{-5mm}\\
	
	\textit{Heterogeneity-induced effects.}  
	Figures~1b-c show that introducing frequency heterogeneity with $\nu = 0.005$ stabilizes previously unstable cyclops and two-cluster states (red markers in Fig.~1a). Supplementary Movies~1 and~2 illustrate the full dynamics of this stabilization process. The mechanism behind this effect can be understood as follows. In the identical oscillator case, the cyclops state requires each oscillator within a cluster to maintain a fixed phase difference ($x$ or $y$) relative to the solitary oscillator. 
	Any deviation from this strict phase alignment may disrupt the cluster structure and destabilize the cyclops state as a whole.
	Frequency heterogeneity, however, introduces small intra-cluster phase variations, broadening the clusters' phase width. While one might expect this to further disrupt the balance required for cyclops formation, the internal phase shifts instead contribute to a compensating effect: oscillators with slightly larger or smaller phase differences balance each other out, allowing the cluster-average phase to satisfy the required condition relative to the solitary node. Thus, the heterogeneity among oscillators directly induces a stable cyclops state (Fig.~1b).
	A similar mechanism applies to the stabilization of two-cluster states, where broadened clusters maintain a constant average phase separation. Unlike the cyclops case, this does not involve a solitary oscillator and is thus less counterintuitive. Strictly speaking, heterogeneity does not stabilize the perfectly symmetric states of the identical oscillator system but instead gives rise to structurally similar configurations with nonzero intra-cluster phase spreads.
	
	\begin{figure}[h!]\center
		\includegraphics[width=0.45\textwidth]{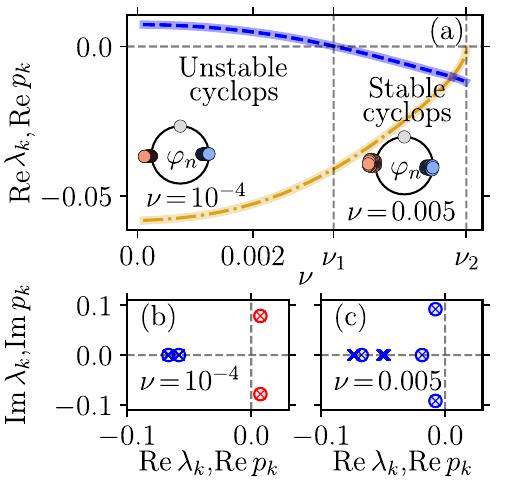}\vspace{-3mm}
		\caption{
			Stabilization of the cyclops state by frequency heterogeneity for the parameter set corresponding to the red circle in Fig.~1a. 
			(a) Real parts of the leading eigenvalues of the full system~\eqref{eq:system_origin_delta_comp1} ($\lambda_k$; dashed and dash-dotted lines) and the reduced model~\eqref{eq:coll_coord} ($p_k$; solid transparent lines). Blue curves: complex-conjugate eigenvalues $\lambda_{1,2}$, $p_{1,2}$ controlling stability; yellow curves: real eigenvalues $\lambda_3$, $p_3$ associated with existence. Stability is achieved for $\nu \in [\nu_1, \nu_2] \approx [0.0034, 0.0057]$, after which the cyclops state disappears. Insets: phase snapshots of the corresponding cyclops states (unstable at $\nu = 10^{-4}$ and stable at $\nu = 0.005$).
			(b, c) Comparison of individual eigenvalues at $\nu = 10^{-4}$ (b) and $\nu = 0.005$ (c); crosses: full model ($\lambda_k$), circles: reduced model ($p_k$). Blue: stable; red: unstable. Only eigenvalues near the imaginary axis are shown.  Parameters: $N = 101$, $\alpha_1 = 1.57$, $\varepsilon_1 = 1.0$.
		}
		\label{fig3}
		\vspace{-5mm}
	\end{figure}
	\begin{figure}[h!]\center
		\includegraphics[width=0.45\textwidth]{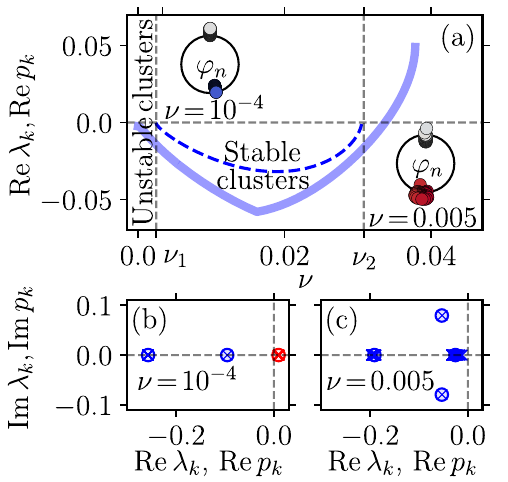}\vspace{-3mm}
		\caption{
			Stabilization of the two-cluster state by frequency heterogeneity for the parameter set corresponding to the red cross in Fig.~1a. 
			(a) Real parts of the leading eigenvalue of the full system~\eqref{eq:system_origin_delta_comp1} ($\lambda_1$; dashed line) and reduced model~\eqref{eq:coll_coord} ($p_1$; solid transparent line). Blue curves: $\lambda_1$ and $p_1$, which govern the stability of the two-cluster state. Stability emerges in the interval $\nu \in [\nu_1, \nu_2] \approx [0.00244, 0.0305]$. Insets: phase snapshots of the corresponding two-cluster regimes (unstable at $\nu = 10^{-4}$ and stable at $\nu = 0.005$).
			(b, c) Comparison of eigenvalues at $\nu = 10^{-4}$ (b) and $\nu = 0.005$ (c); crosses: full model ($\lambda_m$), circles: reduced model ($p_m$). Blue: stable; red: unstable. Only eigenvalues near the imaginary axis are shown.  Parameters: $N = 101$, $\alpha_1 = 1.57$, $\varepsilon_1 = 1.0$.
		}
		\label{fig4}
		\vspace{-5mm}
	\end{figure}
	Figure~\ref{fig3} shows that increasing $\nu$ initially produces an unstable cyclops state with tightly grouped oscillators; further increase stabilizes it over a broad range $\nu \in [\nu_1, \nu_2]$. Similarly, Fig.~\ref{fig4} demonstrates that for the two-cluster regime, small $\nu$ first destroys the unstable two-cluster symmetric state at $\nu = 0$, then induces a misaligned two-cluster configuration, which remains stable over $\nu \in [\mu_1, \nu_2]$.\\
	
	\textit{Reduction via collective coordinates.}  
	To track the evolution of relative phase differences $\varphi_n = \theta_n - \theta_N$ with respect to the solitary oscillator, we transform the original system~\eqref{eq:system_origin} into the phase difference system for $n = 1, \dots, N{-}1$. Using the identity $\sin \phi = \mathrm{Im}[e^{i\phi}]$, we write the phase difference system compactly in complex form:\vspace{-2mm}
	\begin{equation}
		\begin{aligned}
			&m \ddot\varphi_n+\dot\varphi_n=\Delta_n\\&+\! \sum\limits_{q=1}^2 \text{Im}\Big[\frac{\varepsilon_q e^{-i \alpha_q}}{N}\Big(\sum\limits_{k=1}^{N-1}e^{iq\varphi_k}+1\Big)\Big(e^{-iq\varphi_n}-1\Big)\Big],
		\end{aligned}
		\label{eq:system_origin_delta_comp1}
		\vspace{-2mm}
	\end{equation}
	where $\Delta_n = \omega_n - \omega_N$ is the detuning relative to the solitary oscillator. Stable fixed points of Eq.~\eqref{eq:system_origin_delta_comp1} correspond to stable distributions of constant relative phases $\varphi_n$, including cyclops states and two-cluster configurations.
	
	Directly identifying such states in large heterogeneous networks is computationally demanding and often elusive. To address this, we employ a mesoscopic reduction via the collective coordinate approach \cite{cox2006bistable,gottwald2015model,hancock2018model,smith2019chaos,berner2023synchronization,smith2020model,yadav2024disparity,fialkowski2023heterogeneous}. Given uniformly distributed natural frequencies $\omega_n$, we approximate the phase profile within each cluster by a linear ansatz:\vspace{-2mm}
	\begin{equation}
		\begin{aligned}
			&\!\!\varphi_n(t)\!\approx\!\hat{\varphi}_n(t)\!=\!\psi_1(t)\!+\!\chi_1(t)\Delta_n, \quad \!\!\!n\!=\!1,\dots, K,\\
			&\!\!\varphi_n(t)\!\approx\!\hat{\varphi}_n(t)\!=\!\psi_2(t)\!+\!\chi_2(t)\Delta_n, \quad \!\!\!n\!=\!K+1,\dots, 2K,
		\end{aligned}
		\label{eq:ansatz1}
		\vspace{-2mm}
	\end{equation}
	where $\psi_\mu(t)$ represents the average phase of cluster $\mu = 1, 2$, and $\chi_\mu(t)$ describes the phase drift within each cluster. The product $\chi_\mu(t)\Delta_n$ accounts for deviations from the collective phase due to oscillator detuning.  In simple terms, the ansatz expresses each oscillator’s phase as the sum of a macroscopic cluster phase and a frequency-dependent drift. This drift captures the microscopic frequency-induced spread of phases within each cluster, relative to the solitary oscillator, yielding a mesoscopic description of the collective dynamics.
	
	Following the collective coordinate approach \cite{smith2020model}, we substitute the ansatz into
	Eq.~\eqref{eq:system_origin_delta_comp1}, we compute the residual (error vector) between the exact and approximate dynamics. The evolution equations for the collective coordinates are obtained by requiring this residual to be orthogonal to the tangent space of the ansatz manifold, spanned by the gradients $\partial \hat{\varphi}_n / \partial \psi_\mu$ and $\partial \hat{\varphi}_n / \partial \chi_\mu$. This Galerkin projection yields the closed system~\eqref{eq:coll_coord}, governing the evolution of mesoscopic cluster parameters under heterogeneous detuning. Full derivation details are provided in the Supplementary Material. These equations for $\psi_\mu(t)$ and $\chi_\mu(t)$ approximate the macroscopic evolution of cluster-averaged phase positions and internal phase spreads in a finite oscillator network with heterogeneous detuning values $\Delta_n$:\vspace{-3mm}
	\begin{equation}
		\begin{aligned}
			&\!\!\!\!m \ddot\psi_\mu\!+\!\dot\psi_\mu\!=\! \sum\limits_{q=1}^2\text{Im}\Big[\frac{\varepsilon_q e^{\!-i \alpha_q}}{N}\Big(\!1\!+\!K S_1^{(q)}\!e^{iq\psi_1}\!\!\\&\!\!+\!K S_2^{(q)}\!e^{iq\psi_2}\!\Big)\Big(\frac{\sigma_{\mu}S_{\mu}^{(q)^{*}}-\delta_{\mu}J_{\mu}^{(q)^{*}}}{\sigma_{\mu}-\delta_{\mu}^2}e^{-iq\psi_{\mu}}\!-\!1\Big)\Big],\\
			&\!\!\!\!m \ddot\chi_\mu\!+\!\dot\chi_\mu\!=\!1\!+\!\frac{1}{\sigma_{\mu}\!\!+\!\delta_{\mu}^2}\sum\limits_{q=1}^2 \!\text{Im}\Big[\frac{\varepsilon_q e^{-i \alpha_q}}{N}\Big(1\!\\&\!\!+\!K S_1^{(q)}\!e^{iq\psi_1}\!+\!K S_2^{(q)}\!e^{iq\psi_2}\Big)\Big(J_{\mu}^{(q)^{*}}\!\!\!-\!\delta_{\mu}S_{\mu}^{(q)^{*}}\Big)e^{-iq\psi_{\mu}}\Big],
		\end{aligned}
		\vspace{-7mm}
		\label{eq:coll_coord}
	\end{equation}
	\begin{equation}
		\delta_{\mu}=\!\!\!\!\!\!\!\!\!\!\sum\limits_{k=(\mu-1)K+1}^{\mu K}\!\!\!\!\!\!\!\!\Delta_k\Big/K, \quad S_{\mu}^{(q)}=\!\!\!\!\!\!\!\!\!\!\sum\limits_{k=(\mu-1)K+1}^{\mu K}\!\!\!\!\!\!\!\!\!\!e^{iq\chi_\mu \Delta_k}\Big/K,
		\vspace{-7mm}
		\label{eq:delta_S}
	\end{equation}
	\begin{equation}
		\sigma_{\mu}=\!\!\!\!\!\!\!\!\!\!\sum\limits_{k=(\mu-1)K+1}^{\mu K}\!\!\!\!\!\!\!\!\!\!\Delta_k^2\Big/K, \quad J_{\mu}^{(q)}=\!\!\!\!\!\!\!\!\!\!\sum\limits_{k=(\mu-1)K+1}^{\mu K}\!\!\!\!\!\!\!\!\!\!\Delta_k e^{iq\chi_\mu \Delta_k}\Big/K,
		\vspace{-2mm}
		\label{eq:sigma_J}
		\vspace{-1mm}
	\end{equation}
	where $\delta_\mu$ is the average frequency detuning, $\sigma_\mu$ is the second moment of the frequency distribution, $S_\mu^{(q)}$ is the $q$th order parameter of phase deviations from the average collective phase, and $J_\mu^{(q)}$ is the corresponding frequency-weighted order parameter, all computed within the $\mu$th cluster.
	
	Stationary solutions of the reduced system~\eqref{eq:coll_coord}, given by vector  $\pmb{\Gamma}^* = (\psi_1^*, \psi_2^*, \chi_1^*, \chi_2^*)$ can be used to reconstruct cyclops states of the full system~\eqref{eq:system_origin_delta_comp1} via the ansatz \eqref{eq:ansatz1}, offering a computationally efficient way to identify candidate cyclops configurations. As shown in Supplementary Fig.~1, the agreement between the reduced and full models is remarkably close, with deviations on the order of $10^{-3}$ for typical parameter values.
	Once candidate cyclops states are identified via the stationary solutions $\pmb{\Gamma}^*$, their linear stability can be readily assessed. This is done by evaluating the eigenvalues $p_1, \dots, p_8$ of the linearized eight-dimensional system around $\pmb{\Gamma}^*$, derived from Eq.~\eqref{eq:coll_coord} (see Supplementary Material). These eigenvalues determine the local stability of the reduced mesoscopic dynamics and serve as a computationally efficient proxy for the full system, whose Jacobian has $2(N{-}1)$ eigenvalues
	$\lambda_1, \dots, \lambda_{2(N-1)}.$ 
	Figure~\ref{fig3} shows that the eigenvalues $\lambda_k$ of the full system~\eqref{eq:system_origin_delta_comp1} and $p_k$ of the reduced model~\eqref{eq:coll_coord} match remarkably well, confirming the predictive accuracy of the mesoscopic approach. In particular, the blue curves in Fig.~\ref{fig3}a show that the leading complex-conjugate eigenvalue pairs $(\lambda_{1,2})$ and $(p_{1,2})$, which govern the stability of the cyclops state, evolve nearly identically in both models as $\nu$ varies.
	
	The reduced system \eqref{eq:coll_coord} also applies to two-cluster states, which, unlike cyclops states, do not involve a solitary oscillator. In this case, all relative phases can be defined with respect to a common zero reference phase, and the $N$th oscillator is simply included in one of the two clusters.
	While the reduced system \eqref{eq:coll_coord} provides a close match to the two-cluster phase distributions observed in the full system, its eigenvalue predictions are less accurate. Figure~\ref{fig4} shows good agreement near the stability boundaries, slightly below (Fig.~\ref{fig4}b) and just above (Fig.~\ref{fig4}c) the stabilization onset at $\nu_1$ but the correspondence degrades toward the middle of the stable $\nu\in[\nu_1,\nu_2]$ interval.

\begin{figure}[h!]\center
		\includegraphics[width=0.45\textwidth]{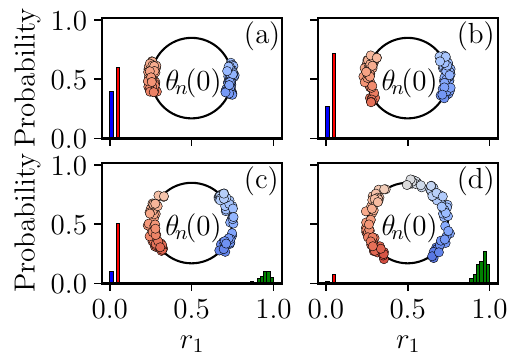}
		\caption{
			Prevalence of heterogeneity-induced cyclops and two-cluster states. Probability of reaching each regime from $100$ randomly initialized two-cluster phase distributions with cluster half-widths $\delta = 0.4$ (a), $\delta = 0.6$ (b), $\delta = 0.8$ (c), and $\delta = 1.0$ (d). Initial phases $\theta_n(0)$ are drawn from two clusters centered at fixed offsets, with examples shown in the insets; initial velocities $\dot{\theta}_n(0)$ are uniformly distributed in $[-0.5, 0.5]$. Blue bins: cyclops states; red: two-cluster states; green: near-synchronous regimes with $r_1 \approx 1$. Cyclops states emerge with probabilities up to 0.4, indicating spontaneous formation of a solitary oscillator from symmetric initial conditions. Parameters correspond to the red circle in Fig.~1a.
		}
		\label{figS2!}
		\vspace{-5mm}
	\end{figure}

	Figure~\ref{figS2!} demonstrates that both cyclops and two-cluster states induced by frequency heterogeneity possess relatively large basins of attraction. When initialized from broad two-cluster phase distributions, the system converges to stationary regimes with high probability up to 70\% for two-cluster states. Remarkably, cyclops states emerge in up to 40\% realizations, despite the absence of a solitary oscillator in the initial conditions. This suggests that the solitary node can self-organize from symmetric initial configurations, highlighting the robustness of the cyclops regime.

	\textit{Conclusions.} 
	We have revealed a counterintuitive effect whereby oscillator frequency heterogeneity can stabilize cyclops and cluster states that are unstable in identical-oscillator networks. This disorder-induced mechanism transforms configurations that are presumably fragile, requiring a precise phase balance, into robust, dynamically accessible regimes across a broad range of initial conditions and heterogeneity sizes. To demonstrate the generality of this effect, we used a simple uniform distribution of oscillator frequencies that requires no fine-tuning to induce stable patterns. However, we expect the mechanism to extend to other unimodal frequency distributions, provided they do not contain excessive outliers. Such outliers can skew the prediction of the mesoscopic reduction and give rise to so-called “rogue” oscillators that detach from clusters \cite{smith2020model}. 
	Our constructive mesoscopic approach that uncovers the hetegenety-induced stabilization mechanism extends the collective coordinate approach by tracking the relative phase offset with respect to the solitary oscillator, which introduces an additional degree of freedom that enables a mesoscopic multi-cluster description. Importantly, the second coupling harmonic gives rise to a slowly evolving second-order parameter that characterizes the cyclops and cluster states, linking their structure to the harmonic interaction and reinforcing the validity of the linear ansatz through a clear separation of time scales. 
	Beyond Kuramoto networks, we expect this disorder-promoted stabilization of cyclops and cluster states to manifest in broader classes of physical and biological systems. In particular, our recent work \cite{munyayev2023cyclops,smirnov2024synaptic} has demonstrated a direct correspondence between second-order Kuramoto models with higher-mode coupling and adaptive networks of canonical theta neurons \cite{ermentrout1986parabolic,gutkin1998dynamics,so2014networks}. This suggests that similar mechanisms may operate in heterogeneous neuronal circuits and other real-world oscillator networks with adaptive or higher-order coupling.\\
	{\it Acknowledgments.}
	This work was supported by the RSF under project  No. 24-72-00105 (to M.I.B.), the MSHE under project No. FSWR-2020-0036 (to V.O.M., L.A.S. and G.V.O.), and the National Science Foundation (USA) under grant CMMI-2009329 and the Georgia State University Brain and Behavior Program (to I.B.).

\bibliography{references_heterogeneity}

\clearpage
\onecolumngrid
\renewcommand{\figurename}{Supplementary Figure}
\setcounter{equation}{0}
\renewcommand\theequation{S.\arabic{equation}}
\setcounter{figure}{0}
\section{Supplementary Material}
\section{Detailed Derivation of the Collective Coordinate Reduction}

This section provides the full derivation of the reduced mesoscopic model introduced in the main text. Our goal is to identify stationary cyclops and two-cluster states in the full heterogeneous oscillator network described by Eq.~(1) of the main paper and construct their approximate representations using a collective coordinate reduction.

The stationary state of system Eq.~(1) for nonidentical oscillators can be conveniently analyzed in a rotating reference frame. We define the relative phases as $\varphi_n = \theta_n - \Omega t$, where $\Omega$ is the common rotational frequency of the system and $n = 1, \dots, N$. We fix the phase of the $N$th (solitary) oscillator as the reference point, choosing $\varphi_N = 0$. 

To simplify notation, we introduce the phase difference relative to the solitary oscillator:
\begin{equation}\label{S1}
	\varphi_n = \theta_n - \theta_N, \quad n = 1, 2, \dots, N-1.
\end{equation}
This substitution effectively eliminates the absolute phase of the solitary node, anchoring the system and reducing the analysis to relative phase dynamics.

Substituting \eqref{S1} into Eq.~(1) in the main text and subtracting the equation for the $N$th oscillator from each of the others yields the following system for $n = 1, 2, \dots, N-1$:
\begin{equation}
	m \ddot\varphi_n+\dot\varphi_n = \omega_n - \omega_N + \sum\limits_{q=1}^{2}\frac{\varepsilon_q}{N} \left[ \sum\limits_{k=1}^{N-1} \left( \sin\big(q(\varphi_k - \varphi_n) - \alpha_q\big) - \sin\big(q\varphi_k - \alpha_q\big) \right) - \sin(q\varphi_n + \alpha_q) + \sin \alpha_q \right].
	\label{eq:system_origin_delta}
\end{equation}
To express the system in terms of intrinsic frequency mismatches, we introduce the detuning parameters:
\begin{equation}
	\Delta_n = \omega_n - \omega_N,
\end{equation}
which quantify the deviation of each oscillator’s natural frequency from that of the solitary node.

To facilitate further analysis, it is convenient to rewrite the system in complex form. Using the identity $\sin(\phi) = \text{Im}\left[e^{i\phi}\right]$, we transform Eq.~\eqref{eq:system_origin_delta} into the following equivalent representation:
\begin{equation}
	m \ddot\varphi_n + \dot\varphi_n = \Delta_n + \sum_{q=1}^2 \text{Im} \left[ \frac{\varepsilon_q e^{-i \alpha_q}}{N} \left( \sum_{k=1}^{N-1} e^{i q \varphi_k} + 1 \right) \left( e^{-i q \varphi_n} - 1 \right) \right],
	\label{eq:system_origin_delta_comp}
\end{equation}
where $n = 1, 2, \dots, N-1$. 

For compactness, we introduce the vector notation
\begin{equation}
	\pmb{\varphi} = (\varphi_1, \varphi_2, \dots, \varphi_{N-1}),
	\label{eq:varphi}
\end{equation}
and define the nonlinear right-hand side as
\begin{equation}
	\Phi_n(\pmb{\varphi}, \Delta_n) = \Delta_n + \sum_{q=1}^2 \text{Im} \left[ \frac{\varepsilon_q e^{-i \alpha_q}}{N} \left( \sum_{k=1}^{N-1} e^{i q \varphi_k} + 1 \right) \left( e^{-i q \varphi_n} - 1 \right) \right].
	\label{eq:Phi}
\end{equation}
This allows the system to be expressed compactly as
\begin{equation}
	m \ddot\varphi_n + \dot\varphi_n = \Phi_n(\pmb{\varphi}, \Delta_n),
	\label{eq:system_origin_delta_comp_vec}
\end{equation}
which we will use as the starting point for the reduction via collective coordinates.

The stationary solutions of Eq.~\eqref{eq:system_origin_delta_comp_vec} correspond to uniformly rotating states of the original system~(1). These solutions are defined by the set
\begin{equation}
	\pmb{\varphi}^{*} = (\varphi_1^*, \varphi_2^*, \dots, \varphi_{N-1}^*),
	\label{eq:stat_sol_phase}
\end{equation}
which satisfy the nonlinear algebraic system
\begin{equation}
	\Phi_n(\pmb{\varphi}^*, \Delta_n) = 0, \quad n = 1, 2, \dots, N-1.
\end{equation}

To analyze the linear stability of the stationary solution~\eqref{eq:stat_sol_phase}, we introduce small perturbations $\delta\varphi_n(t)$ and linearize Eq.~\eqref{eq:system_origin_delta_comp_vec} around $\pmb{\varphi}^*$, which yields the variational equation:
\begin{equation}
	m \delta\ddot\varphi_n + \delta\dot\varphi_n = \sum_{k=1}^{N-1} \left. \frac{\partial \Phi_n}{\partial \varphi_k} \right|_{\pmb{\varphi}^*} \delta\varphi_k, \quad n = 1, 2, \dots, N-1.
	\label{eq:system_origin_delta_comp_vec_lin}
\end{equation}
Assuming solutions of the form $\delta\varphi_n(t) = D_n e^{\lambda t}$, the linearized system reduces to the eigenvalue problem
\begin{equation}
	\pmb{M_P} \, \mathbf{D} = \Lambda \, \mathbf{D},
	\label{eq:ev_problem_phase}
\end{equation}
where $\mathbf{D} = (D_1, D_2, \dots, D_{N-1})^T$, $\Lambda = m \lambda^2 + \lambda$, and $\pmb{M_P} = \left. \frac{\partial \pmb{\Phi}}{\partial \pmb{\varphi}} \right|_{\pmb{\varphi}^*}$ is the Jacobian matrix evaluated at the stationary solution.

The eigenvalues $\Lambda_k$ ($k = 1, \dots, N-1$) obtained from Eq.~\eqref{eq:ev_problem_phase} yield the corresponding set of dynamical eigenvalues $\lambda_k$ ($k = 1, \dots, 2N-2$), which determine the linear stability of the stationary state~\eqref{eq:stat_sol_phase}.
Directly identifying such stationary states to be used in the linear stability analysis in large heterogeneous oscillator networks is computationally demanding and often elusive. 

To overcome this obstacle, we introduce a reduction approach based on the collective coordinate framework \cite{cox2006bistable,gottwald2015model,hancock2018model,smith2019chaos,berner2023synchronization,smith2020model}. This method approximates the phase dynamics of clustered states using a low-dimensional ansatz that captures the essential structure of cyclops and two-cluster configurations. As we demonstrate below, this reduction not only enables efficient computation of candidate solutions but also yields reliable predictions for the stability.

We now apply the collective coordinate method to approximate the dynamics of cyclops states. This approach is based on constructing a liner ansatz $\pmb{\hat{\varphi}}$ for the oscillator phases,
\begin{equation}
	\varphi_n(t) \approx \hat{\varphi}_n(\pmb{\Gamma}(t); \Delta_n), \quad n = 1, 2, \dots, 2K,
	\label{eq:ansatz_main}
\end{equation}
where $\pmb{\Gamma}(t)$ is a vector of collective coordinates that evolve in time and parameterize the effective dynamics of the cyclops state. Given that the natural frequencies $\omega_n$ (and thus the detunings $\Delta_n$) are uniformly distributed, the relative phases $\varphi_n(t)$ within each cluster can be well-approximated by linear functions of the form:
\begin{equation}
	\begin{aligned}
		&\varphi_n(t) \approx \hat{\varphi}_n(t) = \psi_1(t) + \chi_1(t)\Delta_n, \quad n = 1, \dots, K, \\
		&\varphi_n(t) \approx \hat{\varphi}_n(t) = \psi_2(t) + \chi_2(t)\Delta_n, \quad n = K+1, \dots, 2K,
	\end{aligned}
	\label{eq:ansatz}
\end{equation}
where $\psi_{\mu}(t)$ represents the collective phase of the $\mu$th cluster and $\chi_{\mu}(t)$ characterizes the linear phase distortion within the cluster due to frequency heterogeneity. The product $\chi_{\mu}(t)\Delta_n$ thus captures the deviation of the $n$th oscillator's phase from its cluster average $\psi_{\mu}(t)$, with $\mu = 1, 2$ denoting the cluster index.

In this representation, the dynamics of the full phase difference system \eqref{eq:system_origin_delta_comp_vec} are approximated by the evolution of four collective coordinates:
\begin{equation}
	\pmb{\Gamma}(t) = \left( \psi_1(t), \psi_2(t), \chi_1(t), \chi_2(t) \right),
\end{equation}
which together define a mesoscopic description of the cyclops regime.

To derive the reduced equations, we substitute the ansatz~\eqref{eq:ansatz} into the full system~\eqref{eq:system_origin_delta_comp}, which introduces an approximation error. The residual (or error) for each oscillator is given by
\begin{equation}
	\xi_n = m \ddot{\hat{\varphi}}_n + \dot{\hat{\varphi}}_n - \Delta_n - \sum\limits_{q=1}^{Q} \text{Im} \left[ \frac{\varepsilon_q e^{-i \alpha_q}}{N} \left( \sum\limits_{k=1}^{N-1} e^{i q \hat{\varphi}_k} + 1 \right) \left( e^{-i q \hat{\varphi}_n} - 1 \right) \right],
\end{equation}
and the full error vector is
\begin{equation}
	\boldsymbol{\xi} = \left( \xi_1, \xi_2, \dots, \xi_{2K} \right).
\end{equation}

Following the standard collective coordinate approach~\cite{smith2020model}, we minimize the error by requiring that it is orthogonal to the tangent space of the ansatz manifold. This yields the following orthogonality conditions:
\begin{equation}
	\braket{\boldsymbol{\xi}, \frac{\partial \boldsymbol{\hat{\varphi}}}{\partial \psi_1}} = 0, \quad
	\braket{\boldsymbol{\xi}, \frac{\partial \boldsymbol{\hat{\varphi}}}{\partial \psi_2}} = 0, \quad
	\braket{\boldsymbol{\xi}, \frac{\partial \boldsymbol{\hat{\varphi}}}{\partial \chi_1}} = 0, \quad
	\braket{\boldsymbol{\xi}, \frac{\partial \boldsymbol{\hat{\varphi}}}{\partial \chi_2}} = 0,
	\label{eq:ort}
\end{equation}
where $\braket{\cdot, \cdot}$ denotes the Euclidean scalar product.

The corresponding derivatives of the ansatz vector with respect to the collective coordinates are
\begin{equation}
	\frac{\partial \boldsymbol{\hat{\varphi}}}{\partial \psi_1} = \left( \underbrace{1\,\,\,\dots\,\,\,1}_K\,\,\,\underbrace{0\,\,\,\dots\,\,\,0}_K \right)^{T}, \quad
	\frac{\partial \boldsymbol{\hat{\varphi}}}{\partial \psi_2} = \left( \underbrace{0\,\,\,\dots\,\,\,0}_K\,\,\,\underbrace{1\,\,\,\dots\,\,\,1}_K \right)^{T}.
	\label{eq:coll_coord_projection1}
\end{equation}
Substituting into Eq.~\eqref{eq:ort} and evaluating the inner products, we obtain the following coupled equations for the cluster-averaged phases $\psi_\mu(t)$ and phase distortions $\chi_\mu(t)$:
\begin{equation}
	\begin{aligned}
		m \ddot{\psi}_\mu + \dot{\psi}_\mu + \left(m \ddot{\chi}_\mu + \dot{\chi}_\mu - 1 \right) \frac{1}{K} \sum\limits_{k=(\mu - 1)K + 1}^{\mu K} \Delta_k \\
		= \sum\limits_{q=1}^{Q} \text{Im} \Bigg[ \frac{\varepsilon_q e^{-i \alpha_q}}{N} \Big(1 + \sum\limits_{k=1}^{K} e^{i q (\psi_1 + \chi_1 \Delta_k)} + \sum\limits_{k=K+1}^{2K} e^{i q (\psi_2 + \chi_2 \Delta_k)} \Big) \\
		\times \left( \frac{1}{K} \sum\limits_{k=(\mu - 1)K + 1}^{\mu K} e^{-i q (\psi_\mu + \chi_\mu \Delta_k)} - 1 \right) \Bigg],
	\end{aligned}
	\label{eq:psi_chi}
\end{equation}
where $\mu = 1, 2$ indexes the two clusters.
For convenience, we introduce the following notation:
\begin{equation}
	\delta_{\mu} = \frac{1}{K} \sum\limits_{k=(\mu - 1)K + 1}^{\mu K} \Delta_k, \quad 
	S_{\mu}^{(q)} = \frac{1}{K} \sum\limits_{k=(\mu - 1)K + 1}^{\mu K} e^{i q \chi_\mu \Delta_k},
	\label{eq:delta_S}
\end{equation}
where $\delta_{\mu}$ is the average frequency detuning within the $\mu$th cluster, and $S_{\mu}^{(q)}$ is the $q$th-order Kuramoto-type order parameter describing the internal phase distribution in that cluster.
Substituting these expressions into Eq.~\eqref{eq:psi_chi} yields a more compact form for the projected equation onto $\partial \boldsymbol{\hat{\varphi}}/\partial \psi_\mu$:
\begin{equation}
	\begin{aligned}
		m \ddot{\psi}_\mu + \dot{\psi}_\mu + \delta_{\mu} \left( m \ddot{\chi}_\mu + \dot{\chi}_\mu - 1 \right) 
		= \sum\limits_{q=1}^{Q} \text{Im} \Bigg[ \frac{\varepsilon_q e^{-i \alpha_q}}{N} \Big(1 + K S_1^{(q)} e^{i q \psi_1} + K S_2^{(q)} e^{i q \psi_2} \Big) \\
		\times \left( S_{\mu}^{(q)*} e^{-i q \psi_\mu} - 1 \right) \Bigg].
	\end{aligned}
	\label{eq:psi_chi_1}
\end{equation}

To complete the Galerkin projection, we now consider the derivatives of the ansatz with respect to $\chi_1$ and $\chi_2$:
\begin{equation}
	\frac{\partial \boldsymbol{\hat{\varphi}}}{\partial \chi_1} = 
	\left( \Delta_1\,\,\, \dots\,\,\, \Delta_K\,\,\, \underbrace{0\,\,\, \dots\,\,\, 0}_K \right)^{T}, \quad
	\frac{\partial \boldsymbol{\hat{\varphi}}}{\partial \chi_2} = 
	\left( \underbrace{0\,\,\, \dots\,\,\, 0}_K\,\,\, \Delta_{K+1}\,\,\, \dots\,\,\, \Delta_{2K} \right)^{T}.
	\label{eq:coll_coord_projection2}
\end{equation}
We further introduce
\begin{equation}
	\sigma_{\mu} = \frac{1}{K} \sum\limits_{k=(\mu - 1)K + 1}^{\mu K} \Delta_k^2, \quad 
	J_{\mu}^{(q)} = \frac{1}{K} \sum\limits_{k=(\mu - 1)K + 1}^{\mu K} \Delta_k e^{i q \chi_\mu \Delta_k},
	\label{eq:sigma_J}
\end{equation}
where $\sigma_{\mu}$ is the second moment of th frequency distribution in the $\mu$th cluster, and $J_{\mu}^{(q)}$ is a weighted $q$th-order order parameter incorporating the detuning.
Projecting onto $\partial \boldsymbol{\hat{\varphi}}/\partial \chi_\mu$, we obtain the second pair of equations:
\begin{equation}
	\begin{aligned}
		\delta_{\mu} (m \ddot{\psi}_\mu + \dot{\psi}_\mu) 
		+ \sigma_{\mu} (m \ddot{\chi}_\mu + \dot{\chi}_\mu - 1) 
		= \sum\limits_{q=1}^{Q} \text{Im} \Bigg[ \frac{\varepsilon_q e^{-i \alpha_q}}{N} \Big(1 + K S_1^{(q)} e^{i q \psi_1} + K S_2^{(q)} e^{i q \psi_2} \Big) \\
		\times \left( J_{\mu}^{(q)*} e^{-i q \psi_\mu} - \delta_{\mu} \right) \Bigg].
	\end{aligned}
	\label{eq:psi_chi_2}
\end{equation}

Solving Eqs.~\eqref{eq:psi_chi_1} and \eqref{eq:psi_chi_2} simultaneously gives the evolution of the collective coordinates $\psi_\mu(t)$ and $\chi_\mu(t)$. For practical implementation, we rewrite them in a decoupled form as:
\begin{equation}
	\begin{aligned}
		m \ddot{\psi}_\mu + \dot{\psi}_\mu 
		= \sum\limits_{q=1}^{Q} \text{Im} \Bigg[ \frac{\varepsilon_q e^{-i \alpha_q}}{N} 
		\Big( 1 + K S_1^{(q)} e^{i q \psi_1} + K S_2^{(q)} e^{i q \psi_2} \Big)
		\left( \frac{ \sigma_{\mu} S_{\mu}^{(q)*} - \delta_{\mu} J_{\mu}^{(q)*} }{ \sigma_{\mu} - \delta_{\mu}^2 } e^{-i q \psi_\mu} - 1 \right) \Bigg],
	\end{aligned}
	\label{eq:psi}
\end{equation}

\begin{equation}
	\begin{aligned}
		m \ddot{\chi}_\mu + \dot{\chi}_\mu 
		= 1 + \frac{1}{\sigma_{\mu} + \delta_{\mu}^2} \sum\limits_{q=1}^{Q} \text{Im} \Bigg[ \frac{\varepsilon_q e^{-i \alpha_q}}{N} 
		\Big( 1 + K S_1^{(q)} e^{i q \psi_1} + K S_2^{(q)} e^{i q \psi_2} \Big)
		\left( J_{\mu}^{(q)*} - \delta_{\mu} S_{\mu}^{(q)*} \right) e^{-i q \psi_\mu} \Bigg].
	\end{aligned}
	\label{eq:chi}
\end{equation}
Equations~\eqref{eq:psi} and \eqref{eq:chi} define the reduced eight-dimensional system (4) in the main text, governing the collective dynamics of clustered oscillators in the presence of frequency heterogeneity. 
\begin{figure}[h!]
	\centering
	\includegraphics[width=0.5\textwidth]{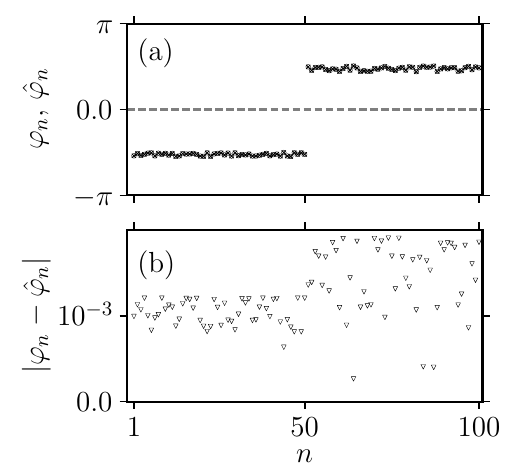}
	\caption{Comparison of stationary phases in the full and reduced systems. (a) Phase values from the full system~\eqref{eq:system_origin_delta_comp_vec} (crosses) and the reduced collective coordinate approximation~\eqref{eq:approx_phase} (circles) based on the solution of Eqs.~\eqref{eq:psi}, \eqref{eq:chi}. (b) Absolute error $|\varphi_n - \hat{\varphi}_n|$ between the exact and approximated phases, which remains below $2 \times 10^{-3}$ for all oscillators. Parameters: $N = 101$, $\alpha_1 = 1.57$, $\varepsilon_1 = 1.0$, $\varepsilon_2 = 0.03$, $\alpha_2 = 0.1$, $\nu = 0.005$.}
	\label{figS1}
\end{figure}
Thus, the stationary solutions
\begin{equation}
	\pmb{\Gamma}^{*} = (\psi_1^*, \psi_2^*, \chi_1^*, \chi_2^*)
	\label{eq:stat_sol_coll}
\end{equation}
satisfying the nonlinear system
\begin{equation}
	\begin{array}{l}
		F_{\mu}(\psi_1, \psi_2, \chi_1, \chi_2) = 0,\\
		G_{\mu}(\psi_1, \psi_2, \chi_1, \chi_2) = 0, \quad \mu = 1, 2
	\end{array}
\end{equation}
define stationary cyclops or two-cluster modes of the reduced system~\eqref{eq:psi},~\eqref{eq:chi}.

These solutions can be used to reconstruct approximate stationary phase configurations of the full difference system \eqref{eq:system_origin_delta_comp_vec}. Specifically, the collective coordinates $\pmb{\Gamma}^*$ yield the approximate phase differences
\begin{align}
	&\hat{\varphi}^*_n = \psi_1^* + \chi_1^* \Delta_n, \quad n = 1, \dots, K,\\
	&\hat{\varphi}^*_n = \psi_2^* + \chi_2^* \Delta_n, \quad n = K + 1, \dots, 2K,
	\label{eq:approx_phase}
\end{align}
which follow directly from the ansatz~\eqref{eq:ansatz}.

Supplementary Figure~\ref{figS1} confirms the accuracy of the reduced model by comparing its predicted stationary phase profile with the full system. The collective coordinate approximation derived from Eqs.~\eqref{eq:psi},~\eqref{eq:chi} closely matches the stationary solution of the full phase difference system \eqref{eq:system_origin_delta_comp_vec} for a cyclops state with $N = 101$, $\alpha_1 = 1.57$, $\varepsilon_1 = 1.0$, $\varepsilon_2 = 0.03$, $\alpha_2 = 0.1$, and $\nu = 0.005$. As shown in Supplementary Fig.~\ref{figS1}b, the maximum deviation between the approximated and full phase values does not exceed $2 \times 10^{-3}$.

To assess the linear stability of the stationary solution~\eqref{eq:stat_sol_coll}, we consider small perturbations $\delta\psi_{\mu}(t)$ and $\delta\chi_{\mu}(t)$ around each collective coordinate and linearize the reduced system \eqref{eq:psi}-\eqref{eq:chi} in its vicinity. This yields the linearized equations
\begin{align}
	&m \delta\ddot\psi_{\mu}+\delta\dot\psi_{\mu}=\sum_{k=1}^{2}\left[\frac{\partial F_{\mu}}{\partial\psi_k} \Bigg|_{\pmb{\Gamma^*}}\!\!\!\!\delta\psi_{k}+\frac{\partial F_{\mu}}{\partial\chi_k} \Bigg|_{\pmb{\Gamma^*}}\!\!\!\!\delta\chi_{k}\right],\\
	&m \delta\ddot\chi_{\mu}+\delta\dot\chi_{\mu}=\sum_{k=1}^{2}\left[\frac{\partial G_{\mu}}{\partial\psi_k} \Bigg|_{\pmb{\Gamma^*}}\!\!\!\!\delta\psi_{k}+\frac{\partial G_{\mu}}{\partial\chi_k} \Bigg|_{\pmb{\Gamma^*}}\!\!\!\!\delta\chi_{k}\right],
	\quad \mu=1,2,
	\label{eq:coll_vec_lin}
\end{align}
where all partial derivatives are evaluated at the stationary point $\pmb{\Gamma^*}$. Substituting exponential perturbations $\delta\psi_{\mu} = B_{\mu} e^{\lambda t}$ and $\delta\chi_{\mu} = C_{\mu} e^{\lambda t}$ reduces the problem to a standard eigenvalue problem:
\begin{equation}
	\pmb{M_C}(B_1,B_2,C_1,C_2)^T = \Lambda(B_1,B_2,C_1,C_2)^T,
	\label{eq:ev_problem_coll}
\end{equation}
where the Jacobian matrix $\pmb{M_C}$ is defined as
\begin{equation}
	\pmb{M_C} =
	\begin{pmatrix}
		\displaystyle\frac{\partial F_{1}}{\partial\psi_1} & \displaystyle\frac{\partial F_{1}}{\partial\psi_2} & \displaystyle\frac{\partial F_{1}}{\partial\chi_1} & \displaystyle\frac{\partial F_{1}}{\partial\chi_2} \\
		\displaystyle\frac{\partial F_{2}}{\partial\psi_1} & \displaystyle\frac{\partial F_{2}}{\partial\psi_2} & \displaystyle\frac{\partial F_{2}}{\partial\chi_1} & \displaystyle\frac{\partial F_{2}}{\partial\chi_2} \\
		\displaystyle\frac{\partial G_{1}}{\partial\psi_1} & \displaystyle\frac{\partial G_{1}}{\partial\psi_2} & \displaystyle\frac{\partial G_{1}}{\partial\chi_1} & \displaystyle\frac{\partial G_{1}}{\partial\chi_2} \\
		\displaystyle\frac{\partial G_{2}}{\partial\psi_1} & \displaystyle\frac{\partial G_{2}}{\partial\psi_2} & \displaystyle\frac{\partial G_{2}}{\partial\chi_1} & \displaystyle\frac{\partial G_{2}}{\partial\chi_2}
	\end{pmatrix}\Bigg|_{\pmb{\Gamma^*}}.
	\label{eq:mc}
\end{equation}
The eigenvalues $\Lambda_k = m \lambda_k^2 + \lambda_k$  ($k=1,2,3,4$) of matrix $\pmb{M_C}$  provide a set of eigenvalues $\lambda_k$ ($k=1,2,\dots,8$) of the stationary mode~\eqref{eq:stat_sol_coll}, which determine its stability.
\end{document}